\documentclass{PoS}

\title{The Very High Energy source catalog at the ASI Science Data Center}

\ShortTitle{GeV and TeV sources catalog at ASI Science Data Center}

\author{\speaker{A.~Carosi}$^{a,b}$, F.~Lucarelli$^{a,b}$, L.~A.~Antonelli$^{a,b}$, P.~Giommi$^{b}$\\

E-mail: \email{alessandro.carosi@oa-roma.inaf.it}\\

{\footnotesize
$^{a}$ INAF, National Institute for Astrophysics, I-00136 Rome, Italy;\\
$^{b}$ Agenzia Spaziale Italiana (ASI) Science Data Center, I-00133 Rome, Italy.\\}
}	

\abstract{
The increasing number of Very High Energy (VHE) sources discovered by the current generation of Cherenkov telescopes made particularly relevant the creation of a dedicated source catalogs as well as the cross-correlation of VHE and lower energy bands data in a multi-wavelength framework. The ``TeGeV Catalog'' hosted at the ASI Science Data Center (ASDC)  is a catalog of VHE sources detected by ground-based Cherenkov detectors. The TeGeVcat collects all the relevant information publicly available about the observed GeV/TeV sources. The catalog contains also information about public light curves while the available spectral data are included in the ASDC SED Builder tool directly accessible from the TeGeV catalog web page. In this contribution we will report a comprehensive description of the catalog and the related tools.
}

\FullConference{The 34th International Cosmic Ray Conference,\\
		30 July- 6 August, 2015\\
		The Hague, The Netherlands}

\begin{document}

\section{Introduction}
\label{sec:intro}

The current wealth of available astronomical data makes particularly important the appropriate handling of such an amount of information in order to achieve the maximum scientific output. Cross correlation and multi-band analysis between data coming from different energy ranges are going to play a key role in modern astrophysics due to the simultaneous availability of new generation space-based and ground-based observatories. An improving multi-wavelength analysis capabilities will be of crucial importance in the VHE band where a new golden-age is approaching thanks to the forthcoming construction of the Cherenkov Telescope Array (CTA). The CTA, a worldwide initiative to build the next generation ground-based VHE gamma-ray instrument will deeply explore our Universe in the VHE gamma-rays and will investigate cosmic non-thermal processes, in close cooperation with both space- and ground-based observatories operating at lower energies of the electromagnetic spectrum, and those using other messengers such as cosmic rays and neutrinos. CTA will serve as an open observatory to a wide astrophysics community and this make particularly relevant the appropriate data handling and data cross matching with other bands.\\

The ASI Science Data Center is involved in the CTA activities in close collaboration with Italian National Institute of Astrophysics (INAF). ASDC is participating to the activities regarding data handling, data processing, data management and data access. In particular, ASDC will be able to adapt its web tools, at the moment used to access data by all the ASDC supported missions, to access also CTA data. ASDC web tools will allow to easy obtain and collect multi-wavelength information on CTA sources. In this context, a new VHE source catalog has been prepared in ASDC and it is available on-line: TeGeV catalog (http://www.asdc.asi.it/tgevcat/). The TeGeVcat is collecting all the information publicly available on the TeV sources observed by Cherenkov detectors including light curves and spectral points.

\section{The ASI Science Data Center}
\label{sec:asdc}

The ASDC is the direct evolution of the first science data center of ASI created to host data of the Italian-Dutch X-ray mission BeppoSaX in late 1990s. Basing on that experience, ASDC has been developed as a facility able to provide high level services and products for data access and analysis to a large scientific community. The ASDC system has been designed to support the activities of several missions. Its strongest peculiarity lies in the cross-correlation of data coming from different archives at different energies.  To this purpose, several tools were developed and are currently available on a web-based interface. The ASDC archive system provide non-simultaneous access to data coming from different (mainly space-based) observatories through an homogeneous HTML interface that constitutes a Multi-Mission Interactive Archive (MMIA). The overall system is driven by a DBMS that involve, depending on the specific mission requirements, the original Browse system or a MySQL relational database.  Fast Visualization of the MMIA data  is provided by a dedicated Multi-Frequency Data Explorer that also points to different internal tools to perform a detailed analysis and the cross-matching between archival catalogs in a multi-wavelength framework. \\

At present, the ASCD hosts data of the most important current high energy astrophysics missions as Fermi, AGILE, NUSTAR, SWIFT and supports at different level other missions like, GAIA and the astroparticle experiments AMS and PAMELA. Data of past experiments like HERSCHEL, PLANCK and BeppoSAX are also hosted while the possibility to perform queries to external catalogs make possible a complete study of the SED of a specific source trough the web-based SED builder tool. 

\section{The TeGeV Catalog}
\label{sec:tegevcat}

\begin{figure}[h]
\centering
\includegraphics[width=1.\textwidth]{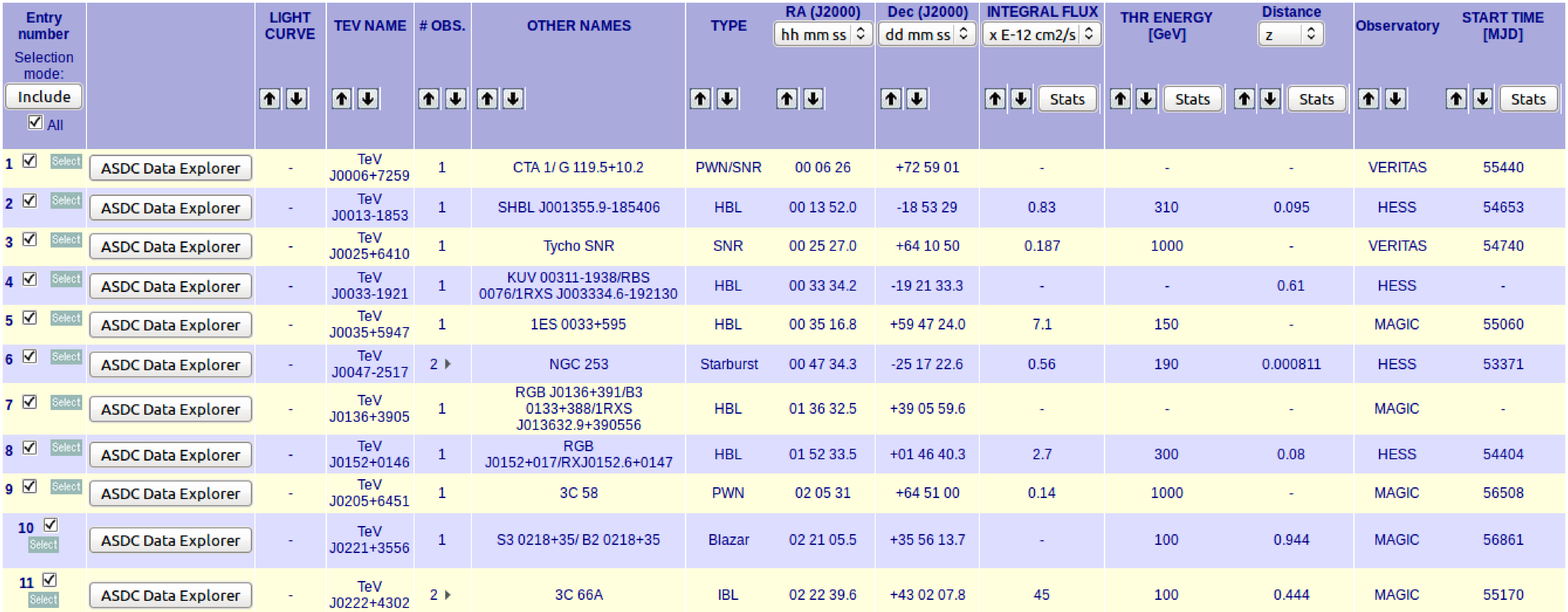}
\caption{Home page of the TeGeV catalog at ASDC servers - http://www.asdc.asi.it/tgevcat/ .}
\label{fig1}
\end{figure}

The TeGeV Catalog @ ASDC is a catalog of VHE sources observed by ground-based Cherenkov telescopes. The TeGeVcat is collecting all the information publicly available about TeV sources observed by the past generation and current generation of imaging Cherenkov telescopes (Fig.~\ref{fig1}). The catalog contains the public light curves data as well as the spectral information that are automatically available in the ASDC SED builder tool (see sec.~\ref{sec:mwl}).

The TeGeV catalog currently contains information of 155 VHE sources divided by class types. Default visualization provides information on source name, coordinates and types. Furthermore, threshold energy ($E_{thr}$), integral flux above $E_{thr}$, distance, name of experiment performing the observation and starting time are provided.\\ 

Several other information are optionally available for the visualization by selecting them at the bottom of the web page. Most of them are again extracted by the literature, like: 

\begin{itemize}
\item the statistical and systematic errors on the TeV centroid position in the sky derived by a 2D-gaussian fit of the TeV excess;
\item the differential flux normalization, the normalization energy of the power law differential spectrum, the power-law spectrum spectral index;
\item source extension (if present) and orientation;
\item observation period and bibliographic references.
\end{itemize}

Other are derived by a further elaboration of the literature data, like:

\begin{itemize}
\item the positional error circle around the TeV source position, calculated by the statistical and systematic errors on the Celestial (or Galactic) coordinates on the TeV excess position.
\item integral flux above 1 TeV, calculated by the spectral information extracted from the literature (when available);
\item information on source flux variability.
\end{itemize}

All these information are displayed following the standard ASDC catalog format. 

\section{Related tools for multi-wavelength analysis}
\label{sec:mwl}

TeGeV catalog is fully implemented within the ASDC web tools that allow different possibilities of data exploring and cross-matching with other energy bands. The multi-frequency \small{DATA EXPLORER} tool, directly available for each source from the catalog web-page, allows the users to explore the sky region around the source position within a customizable radius and showing neighbour sources belonging to different catalog in radio, infrared, optical as well as  X and gamma-rays.\\

\begin{figure}[!t]
\centering
\includegraphics[width=.7\textwidth]{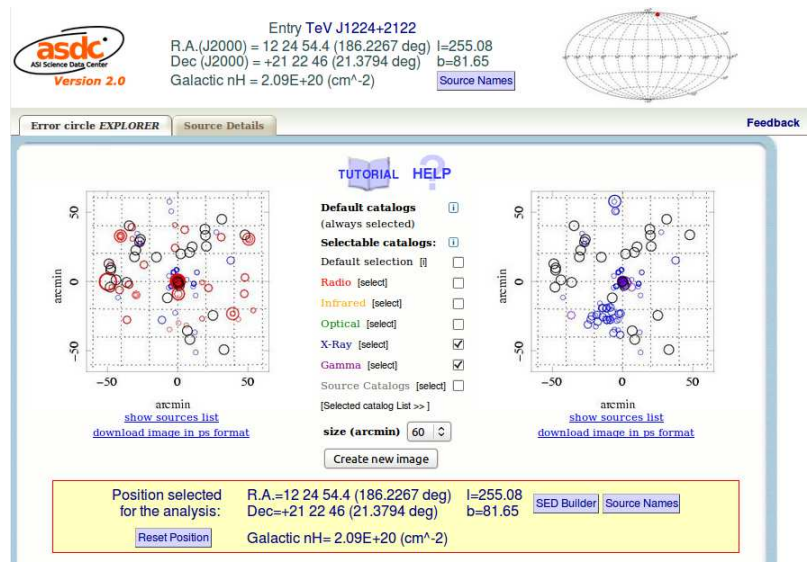}
\caption{The \small{DATA EXPLORER} interface window }
\label{fig2}
\end{figure}

The \small{DATA EXPLORER} (Fig.\ref{fig2}) tool further provide access to several additional services as the possibility to perform query to both internal and external catalogs and archive as well as it provide the access to different web-based data analysis tools. The full source spectral energy distribution (SED) is than directly accessible trough a dedicated \small{SED Builder} tool.\\  

\begin{figure}[!t]
\centering
\includegraphics[width=.85\textwidth]{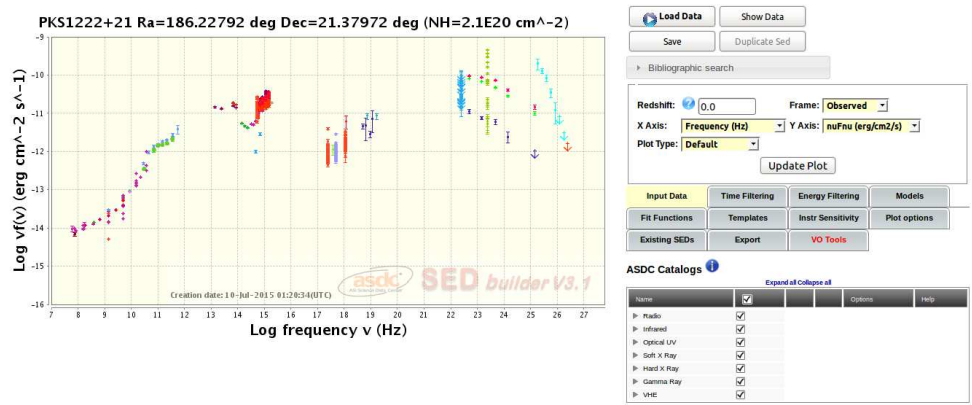}
\caption{The \small{SED BUILDER} tool interface}
\label{fig3}
\end{figure}

The \small{SED Builder} (Fig.\ref{fig3}) is a web based program developed at the ASDC to produce and display the SED of astrophysical sources. The tool combines data from several missions and experiments, both ground and space-based, together with catalogs and archival data. Proprietary data can also be properly handled. Based on a Java code and a MySQL database system, the ASDC \small{SED Builder} provides different functionalities and several plot options for the analysis of the SEDs.




\end{document}